\documentclass[a4paper,11pt]{article}
\title{Numerical study of the transition of the four dimensional Random Filed Ising Model}
\author{Roberto Sacconi \\
        Dipartimento di Fisica, Universit\`a di Roma, ``La Sapienza'', \\
        Piazzale A.Moro 2, 00185 Roma, Italy}

\usepackage{makeidx,showidx}
\usepackage{epsfig}

\begin{document}
\maketitle
\begin{abstract}
We study numerically the region above the critical temperature of
the four dimensional Random Field Ising Model. Using a cluster dynamic
we measure the connected and disconnected magnetic susceptibility and the 
connected and disconnected overlap susceptibility. We use a bimodal
distribution of the field with $ h_R=0.35T $ for all temperatures
and a lattice size $ L=16 $. Through a least-square fit we determine the
critical temperature at which the two susceptibilities diverge. We also 
determine the critical exponents $ \gamma $ and $ \overline{\gamma} $.
We find the magnetic susceptibility and the overlap susceptibility diverge at 
two different temperatures. This
is coherent with the existence of a glassy phase above $ T_c $.
Accordingly with other simulations we find $ \overline{\gamma}=2\gamma $.
In this case we have a scaling theory with two independent critical exponents.
\end{abstract}

\section{Introduction}

In the last few years the Random Field Ising Model \cite{Young}, or RFIM, has
attracted a lot of attention. Despite  great efforts the critical behaviour
of the model is still  not clear.
Both numerical and analytical studies have shown that in three dimensions at 
low temperature and sufficiently small field strength there is a transition
from a disordered phase to a long range ordered phase.
This result was first suggested by Imry and Ma \cite{Imry-Ma}. They consider the possibility that
the model could be split in clusters of dimension $ L $.
Through a direct comparison between ferromagnetic and random field
energy they found a value of two for the lower critical dimension $ d_l $.
Subsequent  arguments \cite{Sourlas}, based on perturbative expansion led to the result 
that the critical behaviour of the RFIM should be  equivalent to that 
of the Ising model in two dimension fewer. This suggested a dimensional 
reduction of two so that, if this should be taken as a general rule,  the lower 
critical dimension should be three instead of two.
However, there is a rigorous proof, see Imbrie \cite{Imbrie}, that  the lower
critical dimension is  two.
It should be shown that, for a certain range of temperatures, the mean field
equation has more than one solution; this is related to the fact that   
this model has a complex free energy landscape.
This is essentially the reason why the dimensional reduction failed.
An accurate numerical investigation of the mean field equation has been done
by Guagnelli {\em et al.} \cite{gumapa} and successively by Lancaster {\em et al.}
 \cite{lamapa}. They have found  that the mean field equation has more than
one solution when the {\em correlation length} is still finite.
In the {\em spin glass} \cite{mepavi} mean field theory we have  a similar 
situation. It seems reasonable to use, in this case too, replica symmetry
breaking theory such as that used by Parisi in that context.
Mez\'ard {\em et al.} \cite{Mez-Young,mezmona}, using RSB techniques and the SCSA approximation \cite{bray}, have
shown the existence of a region above the critical temperature in which
there should be a ``glassy'' phase.
In this case we have two different values of the critical temperature:
one called $ T_c $, which is the usual critical temperature of a ferromagnetic
system, and another , called $ T_b  $ so that $ T_b>T_c $, at which we have a 
transition from  a paramagnetic phase to a ``glassy'' phase.
In section two we first discuss the scaling theory of the model and then we 
introduce the concept of replica susceptibility. In section three we give a 
brief description of the algorithm used and then we show our results and
conclusions in the last two section. 

\section{Theory}
The RFIM is defined by the Hamiltonian
\begin{equation}
        H_{rfim}=-J\sum_{<i,j>}\sigma_{i} \sigma_{j} - \sum_{i} h_{i} \sigma_{i}.
        \label{hamiltoniana}
\end{equation}

The variable $ \sigma_{i} $ are Ising like spin and $ h_{i} $ are independent
random variable with mean $ \left<h_{i}\right>_{av}=0 $ and variance
$ \left<h_{i}^{2}\right>_{av}=h_{R}^{2} $. Typical distribution used is the
Gaussian or  bimodal distribution.

We first discussed the prediction of the scaling theory.
Bray and Moore and independently, Fisher \cite{Bray-Moore,Fisher} have
proposed a scaling theory based on the assumption of a second order phase
transition with a zero temperature fixed point.
At $ T=0 $ for the correlation length, as usual, we expect a power law behaviour given by
\[
        \xi \propto t^{-\nu}.
\]
In this case $ t $ is such that
\[
        t=\frac{h_R}{J}-(\frac{h_R}{J})^{*},
\]
where $ (h_R /J)^{*} $ is the value of $ h_R /J $  at the fixed point.
The other relevant parameters are
\[
\begin{array}{ccc}
      h & , & J.
\end{array}
\]
where $ h $ is an uniform external field.
Because for the RFIM the coupling constant is not fixed this yields to a    change
of the energy scale. In this case we obtain the modified hyperscaling 
relation
\[
        (2-\alpha)=(d-y)\nu,
\]
where $ y $ is the critical exponent related to $ J $ in particular
\[
        J'=b^{y}J.
\]

Another difference with the Ising model is related to the correlation
function. The presence of means over the quenched field causes the 
correlation function to have two different types of behaviour.
We have a connected and a disconnected correlation function.
\begin{equation}
	 G_{con}(r) \equiv \overline{<\sigma_{0}\,\sigma_{r}>-<\sigma_{0}><\sigma_{r}>}=\frac{T}{r^{d-2+\eta}}\,g(r/\xi).
\label{connessa}
\end{equation}
\begin{equation}
 G_{dis}(r) \equiv \overline{<\sigma_{0}><\sigma_{r}>}=\frac{T}{r^{d-4+\overline{\eta}}}\,g^{\prime}(r/\xi) .
\label{sconnessa}
\end{equation}

This defines another critical exponent $ \overline{\eta} $.
The $ \left<\dots\right> $ and the $ \overline{(\dots)} $ denotes, respectively, the thermal average
and the average over the different random field configuration.
The other scaling relations are still valid in this case
\[
\begin{array}{lcll}
 \alpha+2\,\beta+\gamma=2 & , &  \delta=\Delta/\beta & \dots
\end{array}
\]

In summary, we have eleven critical exponents and eight scaling
relation. There seems to be a phase transition with three
independent exponents. Schwartz and Soffer \cite{Schwartz} have
demonstrated the inequality
\begin{equation}
         \overline{\eta} \leq 2\eta.
         \label{disuguaglianza}
\end{equation}

Numerical simulations \cite{Rieger} have suggested that (\ref{disuguaglianza})
should be fulfilled like an equality. In this case we return
to a two independent exponents transition. 

If the transition is a spin glass like transition then the
correct order parameter of the theory is the {\em overlap}
\begin{equation}
        q=\overline{\left<\sigma_{i}\right>^{2}}.
        \label{overlap}
\end{equation}

In this case we are not interested in the magnetisation
correlation function but in the {\em replica correlation function}.
More in detailed we can define a magnetic susceptibility and an
overlap susceptibility
\begin{equation}
\begin{array}{lcl}
   \overline{\chi(m)_{con}} &  = & N[\overline{\langle m^{2} \rangle}-\overline{\langle m \rangle^{2}}] \\
 
   \overline{\chi(m)_{dis}} &  = & N[\overline{\langle m \rangle^{2}}] \\
 
   \overline{\chi(q)_{con}} & = & N[\overline{\langle q^{2}\rangle}-\overline{\langle q \rangle^{2}}] \\
 
   \overline{\chi(q)_{dis}} & = & N[\overline{\langle q \rangle^{2}}]. \\
 \label{suscettive}
\end{array}
\end{equation}
where $ N $ is the volume of the lattice and we have a distinction between the connected and the disconnected parts.
In this paper by means of Monte Carlo simulation we measured the
four quantities in (\ref{suscettive}) in the region slightly above the
critical temperature. In this way we are able to make a comparison
between the susceptibility related to the magnetisation and that
related to the overlap.
If the three temperature transition scheme proposed in \cite{mezmona}
is correct then the overlap and the magnetic susceptibility must diverge at two
different temperatures.

\section{The algorithm}
\label{algoritmo}
The algorithm used to do the simulations  is a generalisation of the
cluster algorithm proposed by Wolff \cite{Wolff} for the Ising Model.
According to the limited cluster flipped algorithm proposed by 
Newmann and Barkema \cite{New-Bark} we have realised an algorithm
capable of flipping more than one spin at a time. The algorithm is
capable to forming clusters with a limited size.
A Monte Carlo step consist of the two following points:
\begin{enumerate}
	\item Build a cluster.

\begin{itemize}
	\item We choose a random site of the lattice. Then we 
choose, according to a certain distribution 
probability\footnote{As is explained in \cite{New-Bark} the appropriate choice 
of this probability distribution is of fundamental importance. In this work
we have used a power law distribution {\em i.e.} $ P(R)=1/R^{\alpha}$ with $
 \alpha=2 $.}, the maximum size, $ R $, of the cluster.
	\item We add similarly oriented neighbouring spins.
If the spin under consideration is  within the allowed distance then we add it
with probability $ 1-\exp{(-2\beta J)} $.
	\item We repeat the above step until there are 
no more spins to add to the cluster.
\end{itemize}
	\item Once the cluster is created we attempt to flip the spins inside it.
\end{enumerate}
The cluster will be flipped with a probability factor
proportional to the random field and to the number $ s $ of 
spins  which might have been added but which are found  just 
outside the radius of the cluster. In detail we have
\[ \left\{ \begin{array}{ll}
	 P_{flip}=\exp{(-2\beta J\,s)}  & \mbox{if}\: m<0 \\
	 P_{flip}=\frac{\exp{(-\beta m)}}{\exp{(\beta m)}}\,\exp{(-2\beta J s)}  & \mbox{if}\: m>0
	\end{array}
\right.  \]
where $ m=\sum_{i\in{C}}{h_{i}\sigma_{i}} $.
As is explained in \cite{New-Bark} this algorithm satisfies the conditions of
ergodicity and detailed balance for the random field model.
A Monte Carlo sweep is obtained  when we have attempted to flip a number of clusters like the volume of the system.
 
We believe that for models
such as the RFIM, this kind of dynamics is capable of strongly reducing the problem
related to the dynamic slowing down approaching the critical
temperature.
Using a single spin flip dynamic  a new configuration is
obtained when we 
try to flip all the spins in the lattice. The probability of flipping
a spin depends on the local fields through a term proportional to
$ p_{flip}(h)=\exp (-2\beta h_{i}) $. It is possible that some spins  are aligned to a large local  field; in this case such
spins are almost impossible to be flipped. Sometimes if we flip this ``pinned'' spin, the configuration obtained should be more
probable than it was. When one of this spins is taken as a part of a cluster
 the effect of the field of such site should be rounded by the other fields 
in the cluster. In this case the procedure that realises
the Monte Carlo dynamics is much more complicated than that
of the Metropolis algorithm. Moreover, this kind of dynamics
depends much from the contest. It is therefore necessary to spend
a lot of time in optimising the algorithm.

\section{Numerical results}
Rieger and Young \cite{Rieger-You,Rieger} have carried out  the most extensive Monte Carlo simulations in three dimensions in order to test the 
scaling relation validity. Using finite size 
scaling techniques they calculate all the critical exponents
both with a Gaussian and with a bimodal probability distribution.

Making use of the cluster algorithm described in section  \ref{algoritmo}
we carried out Monte Carlo simulation in order to search for
numerical evidence of the existence of a ``spin glass'' like
phase transition in the region above the critical temperature for a four dimensional
lattice. 
 To this end, at each Monte Carlo sweep, for each disorder
realisation we measured the average magnetisation $\left< m \right>$ its
square $\left< m^{2} \right>$, the average overlap $ \left< q \right> $ and its square
 $\left< q^{2} \right>$; with $ m=1/N\sum_{i=1}^{N}\sigma_{i} $ and 
$ q=1/N\sum_{i=1}^{N}\sigma_i \tau_i$ where $\sigma_i $ and $\tau_i$ are two
generic spins of two replica of the system.
In this way we can calculate the four quantities given in (\ref{suscettive}).
The two replica are such that they have the same
realisation of the disorder. They will approach
the equilibrium following two different markovian processes, in this case 
they can be considered independent.
We performed the calculation using a four dimensional lattice
of size $ L=16 $ with periodic boundary conditions.
We have measured the quantities in (\ref{suscettive}) for $ 13 $ different values
of the temperature for each realisation of the disorder.
Starting from a high temperature region we have cooled the 
system until  ten percent of the critical temperature.
The hardest region to simulate is the one near the critical temperature. For these temperatures the system takes a great deal of
time to reach equilibrium. Near $ T_{c} $ up to $ 250 $ Monte Carlo
sweeps  were needed to balance the system. 
At temperatures $ T $ sufficiently greater than $ T_c $ the system balances
very fast. In contrast, near $ T_c $ the time needed to balance the 
system became too much long. For this reason if we took the same
amount of Monte Carlo sweeps at each temperature we would waste time.
We have used a range of temperature varying from $ 8.33 $ to $ 5.97 $
that correspond to $ 0.12 \leq \beta \leq 0.1675 $ where $ \beta=1/T $.

Starting from $ \beta=0.12 $ we have cooled the system through the
following rules
\begin{eqnarray}
	\beta & = & 0.12+(k-1)0.005, \label{beta1} \\
	      &   &  k=1 \ldots 8      \nonumber \\ 
	\beta & = &  0.1575+(k-1)0.0025,  \label{beta2} \\
	      &   &  k=9 \ldots 13. \nonumber
\end{eqnarray}
In this way we have more points near $ T_c $.
At each temperature we have used a Monte Carlo sweep number ($\#MCs$) given by
\begin{eqnarray*}
	\mbox{\#}MCs & = & 64 k^2   \\
                       &   &  k=1 \ldots 13 
\end{eqnarray*}

A third of these have been used to balance the system and the rest
to take the measurements. For  $ k=13 $, {\em i.e.} near $ T_c $, from the 
$10816 \; \#MCs,$
the first $ 3605 $ has been used to reach the equilibrium.
This is an order greater than the calculated equilibration time.
In this way, in the hardest region to simulate the thermal average there have 
been 
performed over more than $ 500 $ uncorrelated measurements.
The average over the disorder has been performed on $ 850 $ samples. 
We need such a large number of samples because the quantities 
in (\ref{suscettive}) are highly non-self-averaging. For this reason the
error caused by the disordered average is an order greater than that given
by the thermal average. The random field has been chosen so that  
$ h_R=0.35T $ for different temperatures.
As was pointed out in \cite{Rieger-You} for greater values of $ h_R/T $ 
the system is too difficult to balance and when the ratio is too small  
the system degenerates in the ferromagnetic model.

The quantity (\ref{suscettive}), near $ T_c $ are well fitted by some power
law of the temperature. Both for the the magnetic susceptibiliy and the
overlap susceptibility we are sufficiently far from $ T_c $ so we can 
neglect finite size effect.
For the connected and disconnected magnetic susceptibility  we used the
following power law
\begin{equation}
\begin{array}{ccc}
	\overline{\chi(M)}      & = & C(T-T_c)^{-\gamma}    \\
	\overline{\chi(M)_{dis}} & = & C_{1}(T-T_c)^{-\overline{\gamma}}. 
\end{array}
\end{equation}

\begin{figure}[h]
\begin{center}
\epsfig{figure=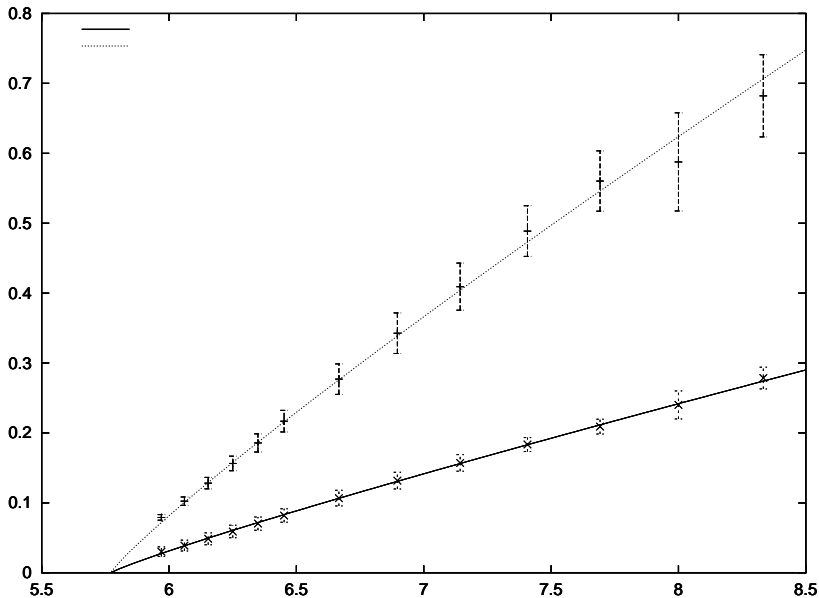,height=8cm}
\caption{\footnotesize \it The full line and the dotted line represent the least-square fit results. The full line is the plot of $ 1/f(x) $ where $ f(x)=8.9(T-T_c)^{-0.94} $ and the dotted line is the plot of \( 1/  f_1(x)^{(-1/2)} \)
 where 
$ f_{1}(x)=12.1(T-T_c)^{-1.91} $.
The intersection with the abscissa 
gives the critical temperature.}
\label{figur1}
\end{center}
\end{figure}

In figures (\ref{figur1}) $ 1/\chi(m)_{con} $ and $ 1/\sqrt{\chi(m)_{dis}} $ 
are 
plotted against the temperature. It is clear from the figure that the two 
exponents $ \gamma $ and $ \overline{\gamma} $ are respectively slightly lower
than one and two. 
 The results of the least-square fit are such that
\begin{equation}
	\begin{array}{ccc}
          T_c=5.72\pm0.04 & \gamma=0.94\pm0.02 & \overline{\gamma}=1.91\pm0.08.
	\end{array}
	\label{result1}
\end{equation}

The jacknife technique is used for the errors; the data are correlated since
the same set of random fields is used at each temperature.
According 
previous simulation \cite{Rieger} the Schwartz inequality \cite{Schwartz} seems
to be valid as an equality. Within the error bars we have $ 2\gamma=\overline{\gamma}$
that is
\[
\overline{\eta}=2\eta.
\]

Considering that $ \overline{<Q^{2}>}\rightarrow 1 $ when $ T \rightarrow \infty $
we expect the connected overlap susceptibility  to have a power behaviour
given by
\begin{equation}
\begin{array}{ccc}
        \overline{\chi(Q)_{con}} & = &  B(T-T_b)^{-\omega}+D  
 \label{fit}
\end{array}
\end{equation}

Because of the presence of the random field also the disconnected overlap 
susceptibility  has  a non vanishing term.
 We expect a power law behaviour given by

\begin{equation}
\begin{array}{ccc}
       \overline{\chi(Q)_{dis}} & = & B_1(T-T_b)^{-\overline{\omega}}+D_1.
	\label{fit2} 
\end{array}
\end{equation}

\begin{figure}[h]
\begin{center}
\epsfig{figure=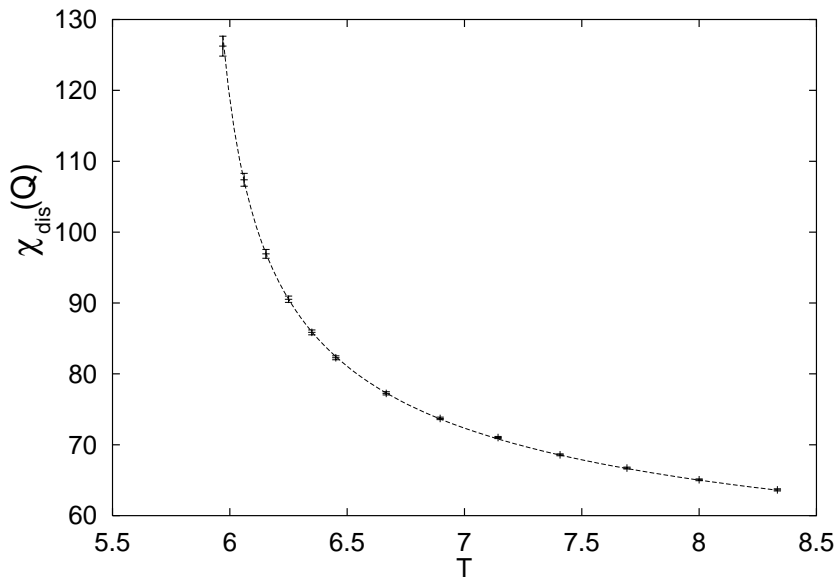,height=8cm}
\caption{\footnotesize \it Disconnected overlap susceptibility. The full line 
is the least-square fit result $f(x)=33(T-5.88)^{-0.42.}+41.$}
\label{figur2}
\end{center}
\end{figure}

We have reported in figure (\ref{figur2}) the least-square fit result. As for
the results in (\ref{result1}) the errors are calculated with the jacknife 
technique. 
 We find
\begin{equation}
	\begin{array}{ccc}
              T_b=5.88\pm0.04 & \omega=0.6\pm0.1 & \overline{\omega}=0.42\pm0.05
	      \label{result2}
	\end{array}
\end{equation}

The (\ref{fit2}) is valid near the critical temperature.
It can be argued that the results found for the two critical temperatures, may
be an artefact of the power-law behaviour used in (\ref{fit2}). There could
be a temperature drift in the non-singular term as the temperature is not too
near  from $ T_c $. To take
care of this effect we can add in (\ref{fit2}) a linear term in the temperature
vanishing near $T_c$. We can use a temperature dependence given by,

\begin{equation}
\begin{array}{ccc}
       \overline{\chi(Q)_{dis}} & = & B_1(T-T^{*})^{-\overline{\omega}}+D_1[1+D_0 \frac{T-T^{*}}{T^{*}}].
	\label{fit3} 
\end{array}
\end{equation}

If we fix $ T^{*} $ we can perform a four parameters fit. When $T^{*}=T_c$ the 
values of $ \chi^{2} $ obtained is greater than that obtained in the previous 
fit. If we set $ T^{*}=T_b$ we recover the results obtained using (\ref{fit2}). In this case the temperature can be neglected and the results are well
represented by the power law given in (\ref{fit2})
The least-square fit results are reported in tabel \ref{tabella}. 

\begin{table}[h]
\begin{center}
\begin{tabular}{|c|c|c|c|} \hline
      &  (\ref{fit2})  & $T^{*}=T_c$    &    $T^{*}=T_b$  \\  \hline \hline 
$ B_1  $       & $ 33\pm 2 $    &   $13.3\pm0.9$&    $32\pm 2 $   \\
$\overline{\gamma}$& $ 0.42\pm 0.02$&   $1.09\pm0.04$& $0.43\pm 0.02$ \\
 $ D_1 $          &  $41\pm 2 $  &   $65\pm 1 $    &  $ 43\pm 2 $      \\
$ D_0  $    & $ 0 $  &   $-0.22\pm0.02 $    &   $-0.03\pm 0.02$   \\ 
$\chi^{2} $  &  $2.14 $  & $4.76$  &     $1.98$      \\   \hline

\end{tabular}

\caption{ \footnotesize \it In the first column we have the least-square fit results obtained using the power law given in (\ref{fit2}). The results in colunm two and three are obtained using the temperature dependence given in (\ref{fit3}). In the last raw we have the values of $\chi^{2}$ calculated during the fit. The values in culomn one and three are almost equal. When $ T^{*}=T_b $ we have $D_{0}\sim0$ so that we can assume that (\ref{fit2}) is a good approximation for all the temperatures used.}
\label{tabella}
\end{center}
\end{table}

\section{Conclusion}
From the data analysis the overlap susceptibility and the magnetic 
susceptibility
seems to diverge at two different points. It turns out that $ T_b>T_c $.
If this is the case the three transition
scheme, obtained through RSB techniques, should be correct.
A more extensive Monte Carlo simulations should be done near $ T_c $, using
finite size scaling techniques, in
order to confirm the result obtained. It should be interesting as well to 
studing the overlap distribution probability in the region under $ T_b $.
Another result is given from the comparison between $ \gamma $ 
and $\overline{\gamma}$.
We find $ \overline{\gamma}=2\gamma $, according to this result we have found
one more scaling relation so that the independent critical exponents
are two instead of three. In four dimension as well, the dimensional reduction
is far from giving the correct result, in fact $ \gamma=0.94 $ is very far 
from $ 7/4 $ which is a prediction of dimensional reduction.
\section*{Acknowledgement}
The author is grateful to G.Parisi and J.J.Ruiz-Lorenzo for useful discussions and suggestions. A special thanks to G.Parisi for his patient and his  disponibility during all over the work.

\end{document}